\documentclass[aps,prl,twocolumn,showpacs,preprintnumbers,amsmath,amssymb,floatfix]{revtex4-1}

\usepackage{pstricks}
\usepackage{graphicx}
\usepackage{dcolumn}
\usepackage{bm}

\newcommand\be{\begin{equation}}
\newcommand\ee{\end{equation}}
\newcommand\bea{\begin{eqnarray}}
\newcommand\eea{\end{eqnarray}}
\newcommand\ba{\begin{array}}
\newcommand\ea{\end{array}}

\begin{document}

\title{Cosmological Signatures of a UV-Conformal Standard Model}

\author{Glauber~C.~Dorsch$^{1}$, Stephan~J.~Huber$^{1}$ and Jose Miguel No$^{1}$\vspace{1mm}}
\affiliation{$^{1}$Department of Physics and Astronomy, University of Sussex, BN1 9QH Brighton, UK}

\date{\today}

\begin{abstract}Quantum scale invariance in the UV has been recently advocated as an attractive way of solving the gauge hierarchy problem arising in the Standard Model.
We explore the cosmological signatures at the electroweak scale when the breaking of scale invariance originates from a hidden sector and is mediated to the Standard Model by gauge interactions (Gauge Mediation). These scenarios, while being hard to distinguish from the Standard Model at LHC, can give rise to a strong electroweak phase transition
leading to the generation of a large stochastic gravitational wave background in possible reach of future space-based detectors such as eLISA and BBO. This relic would be the cosmological imprint of the breaking of scale invariance in Nature.
\end{abstract}

\maketitle

In the Standard Model (SM), the fact that spontaneous breaking of electroweak (EW) symmetry is driven by a fundamental scalar leads to a puzzle concerning the naturalness of the theory. On the one hand, this scalar should have a mass of the order of the EW scale $v = 246$~GeV, recently confirmed by the LHC discovery of a particle consistent with the SM Higgs boson and with mass $m_h \sim 125$~GeV~\cite{Aad:2012tfa, Chatrchyan:2012ufa}. On the other hand, no symmetry in the SM protects $m^2_h$ from receiving quantum corrections scaling as $\delta m_h^2 \sim M^2$ for every energy scale $M$ to which the Higgs is sensitive, so one would expect $m_h$ to be of the order of the largest energy scale at which some new physics enters, e.g. $M_{\mathrm{Pl}}$. This ``{\it gauge hierarchy problem}" 
hints at the existence of some symmetry at energies above the EW scale which forbids or suppresses the large $\delta m_h^2$ contributions altogether.

In that respect, the idea that Nature may be exactly scale invariance at high energies (in the UV) \cite{Salam:1970qk} as an explanation for the lightness of the Higgs 
\cite{Bardeen:1995kv} has recently attracted renewed interest, see {\it e.g.} \cite{Tavares:2013dga,Tamarit:2013vda,Antipin:2013exa,Abel:2013mya} (exact UV scale invariance has also been widely discussed in the context of Asymptotic Safety, see \cite{Litim:2011cp} for a recent review).
Scale invariance would have to be broken at some energy scale $f_c$, above which threshold all quantum corrections to $m_h$ would be forbidden by the symmetry. The mass of the Higgs being sensitive to $f_c$ \cite{Tavares:2013dga}, this new scale should be of order of the TeV energy scale.
However, it has been recently shown \cite{Abel:2013mya} that when the breaking of scale invariance originates in a hidden sector and is mediated to the SM sector via gauge interactions (in analogy to Gauge Mediation in the context of supersymmetry breaking theories), $f_c$ could be significantly higher. Moreover, 
gauge mediation of exact scale breaking (GMESB) allows to compute the effect of breaking of scale invariance on the SM, under general assumptions about the properties of the hidden sector. The Higgs mass $m_h$ would then vanish at tree-level and 
emerge from the breaking of scale invariance via loop-corrections involving the SM gauge bosons, naturally explaining the hierarchy $v \ll f_c$ \cite{Abel:2013mya}. Due to this hierarchy, GMESB scenarios may in principle be very hard to probe at LHC.

In this letter we show that there are very important differences between the SM and these scenarios from the point of view of electroweak cosmology. 
In the SM the electroweak phase transition (EWPT) is known to be a smooth cross-over~\cite{Kajantie:1996mn, Karsch:1996yh, Csikor:1998eu}, leaving no trace in the early universe. In contrast, GMESB scenarios generically predict a strongly first order EWPT, due to the form of the scalar potential arising from the breaking of scale invariance.
For a considerable fraction of the parameter space, the EWPT gives rise to a large stochastic background of gravitational waves, within reach of planned and future space-based gravitational wave observatories such as eLISA or BBO. Moreover, the requirement of a consistent cosmological evolution allows us to obtain information on $f_c$ and the breaking of scale invariance. Altogether, it is likely that breaking of scale invariance would leave an observable imprint in the Early Universe.

The letter is organized as follows: In section I we discuss the form of the radiatively generated Higgs potential in GMESB scenarios, as described in \cite{Abel:2013mya}. 
In section II we analyze the nature and dynamics of the EWPT, and in section III we explore a potentially observable gravitational wave background as a smoking-gun signature of these models, together with the implications of a viable cosmological evolution. In section IV we discuss the interplay between cosmological and collider probes of these scenarios.
We conclude in section V.

\vspace{-5mm}

\subsection{I. Higgs Potential from Quantum Scale Invariance}

\vspace{-3mm}

We start with a scenario in which scale invariance is broken in a ``hidden sector'', interacting with the SM only via SM gauge interactions. GMESB  
implies that the only terms in the Lagrangian mixing the hidden and visible sectors take the form
\be 
\mathcal{L} \supset g A_{\mu}^a \left(J_{\rm vis}^{\mu a} + J_{\rm hid}^{\mu a} \right)
\ee
where $a$ is a gauge group index and $J_{\rm vis}, J_{\rm hid}$ are currents in the visible and hidden sectors, respectively.

The full Lagrangian obeys scale invariance only if the tree-level Higgs mass vanishes, so the only free parameter in the Higgs potential is a quartic coupling $\lambda$. A mass is however generated by loop corrections once scale invariance is broken. The relevant contributions must be at least of 2-loop order: one via which scale breaking is mediated to the SM (through corrections to the gauge boson propagators), and another loop coupling these gauge bosons to the Higgs scalar (additional loops are of course possible, {\it e.g.} including the scalar itself, see \cite{Abel:2013mya}). Due to this 2-loop suppression we have $m_h\ll f_c$, with $f_c \sim \mathcal{O}(1-100$) TeV. The resulting effective potential can be written as \cite{Abel:2013mya}
\be 
	V_{\rm eff}=\frac{3}{2}\,\textrm{Tr}\int \frac{d^4p}{(2\pi)^4}\log\left(p^2 + m_V^2 + g^2C_{\rm vis} + g^2C_{\rm hid}\right)
	\label{V_eff}
\ee
with the trace over the gauge group indices, $m_V$ the gauge boson mass and $C_{\rm vis},~C_{\rm hid}$ parametrizing the corrections to the two-point functions of $J_{\rm vis}$ and $J_{\rm hid}$, respectively. GMESB ensures that all contributions to the Higgs potential involve $f_c$, so we restrict ourselves to
\be 
\delta V_{\rm eff} \equiv \left. V_{\rm eff}-V_{\rm eff}\right|_{f_c=0}. 
\ee
Each term in the difference has divergences that do not depend on $f_c$, while $\delta V_{\rm eff}$ is well-defined.

\begin{figure}
	\includegraphics[width=0.4\textwidth, clip]{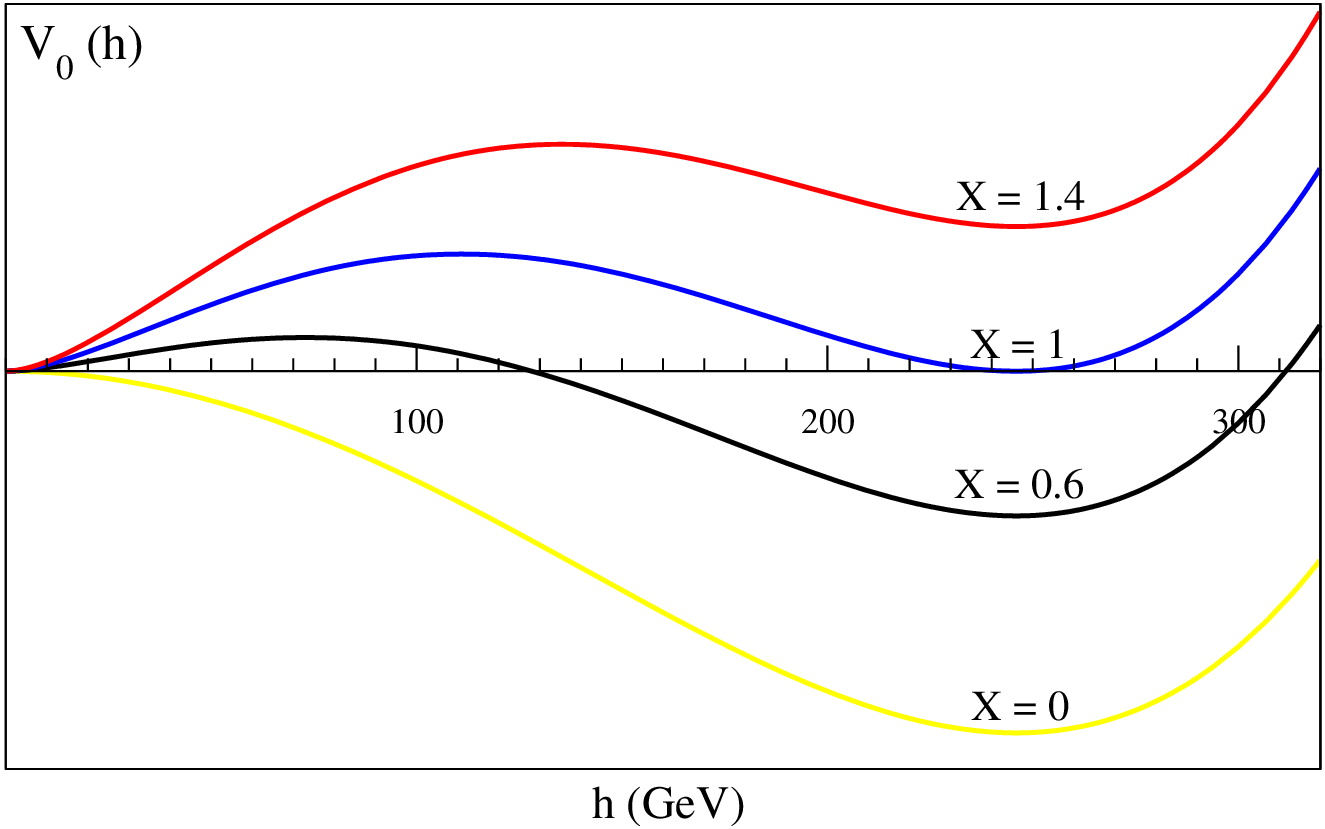}
	\caption{The Higgs effective potential $V_0$ for $X=0, 0.6, 1, 1.4$.}
	\label{fig:Potential}
\end{figure}

A major advantage of GMESB scenarios is that we need not know the details of $C_{\rm hid}$ to compute the Higgs effective potential, whereas the $C_{\rm vis}$ take the same form as in the SM. Following \cite{Abel:2013mya}, one can expand $\delta V_{\rm eff}$ in powers of  the gauge couplings, keeping only the dominant, large momentum ($p^2 \gg v^2$) contributions of the integral in (\ref{V_eff}), for which case a simple representation of $C_{\rm vis}$ exists. The coefficients of the resulting potential will depend on the details of the hidden sector, but requiring  the electroweak minimum $v$ and Higgs mass $m_h$ resulting from it to have the correct value, one finally arrives at \cite{Abel:2013mya}
\be
	\delta V_{\rm eff} \equiv V_0 =  - \frac{m_h^2}{4}\, h^2 \left(1 + X\log\left[\frac{h^2}{v^2}\right]  \right) +
		    \frac{\lambda}{4}\, h^4
	\label{Higgs_potential}
\ee
where $X\equiv \frac{2 v^2 \lambda}{m_h^2}-1$. The details of the breaking of scale invariance are thus encoded in the value of $X$ (or, alternatively, of $\lambda$). The SM case is recovered for $X = 0$. In general, however, the logarithmic term gives a positive contribution to the potential (in the region $h<v$) for $X>0$, giving rise to a 
potential barrier between the EW symmetric and EW broken minima even at zero temperature, as shown in Fig.~\ref{fig:Potential}. 
As discussed in \cite{Abel:2013mya}, weakly coupled realizations of the hidden sector (with the assumption that the SM gauge interactions do not unify at higher energies) precisely yield $X>0$. Note that for $X > 1$ the symmetric phase is the global minimum of the potential (see Fig. 1). As discussed in the next section, this leads to an inconsistent thermal history of the universe, since EW symmetry would never be broken.

\subsection{II. The Electroweak Phase Transition}

\vspace{-3mm}

At finite temperature the Higgs field is surrounded by a thermal plasma of particles. 
As a result, the free-energy of the system is minimized not when the Higgs field is at the minimum of $V_0$, but of an effective thermal potential $V\equiv V_0+V_T$. 
The general expression for $V_T$ at 1-loop \cite{Dolan:1973qd} (including resummed contributions from bosonic ``{\it ring}" diagrams \cite{Carrington:1991hz}) is given by
\bea
	V_T =  \frac{T^4}{2\pi^2}\sum_{a} N_a \int_0^{\infty} dx\, x^2 \, \log\left[1 \pm e^{-\sqrt{x^2+\frac{m^2_a(h)}{T^2}}}\right] \nonumber \\
	+ \frac{T}{12\pi}\sum_{b} \bar{N}_b \left(m^3_b(h) - [m^2_b(h)+\Pi^2_b(T)]^{3/2} \right)\,\,\,
	\label{V_T}
\eea
where in the first term we sum over particles coupling to the Higgs, with numbers of degrees of freedom $N_a$ and a $-$ ($+$) sign for bosons (fermions). 
The dominant contributions are given by $a=W, Z,t$, for which $N_a=(6,3,-12)$. Contributions from the rest of quarks and leptons are negligible, while those from the Higgs and Goldstone bosons are subdominant and can be safely ignored. The second term in (\ref{V_T}) corresponds to the bosonic {\it ring} contributions, to which the longitudinal component of gauge bosons contribute ($\bar{N}_b=(2,1)$ for $b=W, Z$). The field-dependent squared masses $m^2_a(h)$ and the thermal squared masses $\Pi^2_b(T)$ are given by
$m^2_W(h) =  g^2 \,h^2/4$, $m^2_Z(h) = (g^2 + g'^2)\,h^2/4$, $m^2_t(h) = y_t^2\, h^2/2$ and $\Pi^2_W = \Pi^2_Z = 11\,g^2 \,T^2/6$.
%
%
The hidden sector particles do not contribute to $V_T$, since their masses are expected to be $m_{\rm hid}\sim f_c\gg v$, so their presence in the plasma during the EWPT (when $T\sim v$) is strongly Boltzmann suppressed (also, due to GMESB, their contributions would only appear at 2-loop level).

We are now ready to analyze the dynamics of the EWPT. For $T \gg v$ the only minimum of $V$ is at $\langle h\rangle (T) = 0$ and EW symmetry is restored. 
As the Universe expands, temperature decreases and a new local minimum develops away from the origin. As discussed above, if $X\ge 1$ this second minimum will never be energetically favoured over the symmetric one and EW symmetry breaking will never take place. 
For $X<1$, a critical temperature $T_c$ exists at which $V$ has two degenerate minima, and for $T < T_c$ it is possible for the Higgs field to tunnel to the broken phase. The rate per unit time and volume of this tunneling process at a given temperature $T$ is given by~\cite{Coleman:1977py,Linde:1980tt}
\be
	\Gamma \sim T^4 \,e^{-F_c/T}.
	\label{nucleation_rate}
\ee
Here $F_c$ is the free-energy of a critical bubble of true vacuum, {\it i.e.} a bubble just large enough for its internal pressure to overcome its surface tension and grow. At $T_c$, a critical bubble has infinite size, so that $F_c \rightarrow \infty$, and the phase transition does not proceed. The nucleation temperature $T_n$ is defined as that for which the nucleation probability of a critical bubble within a Hubble volume approaches unity. This happens when $F_c/T \approx 140$, which sets the temperature at which the EWPT effectively starts (for large supercooling, a more accurate procedure is needed to determine $T_n$, see {\it e.g.} \cite{Huber:2013kj}). 
%
%
%
%
%
\begin{figure}[ht]
\begin{center}
	\includegraphics[width=0.45\textwidth, clip ]{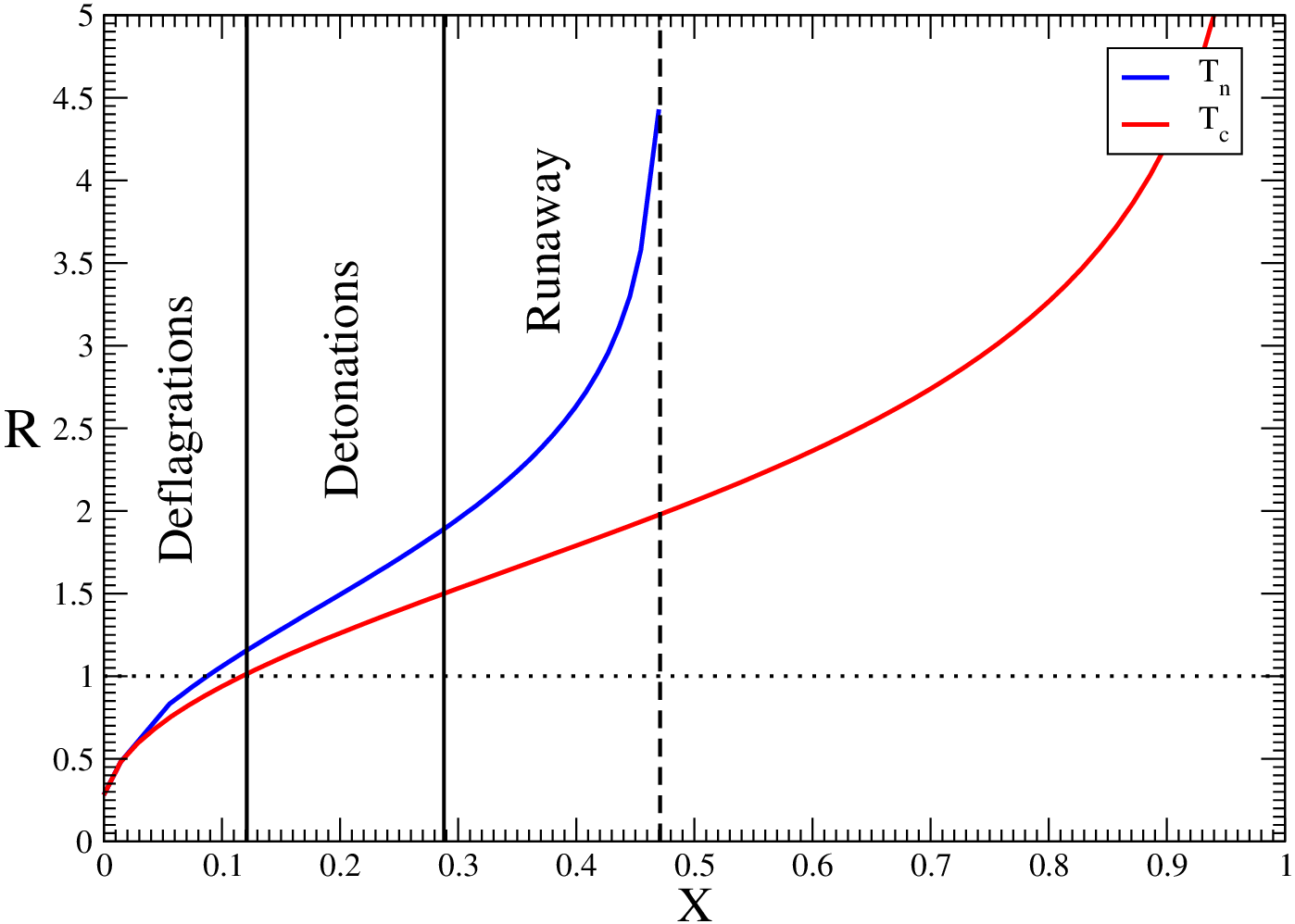}
	\caption{\small Phase transition strength computed at critical (red) and nucleation (blue) temperatures. The dotted horizontal line marks the limit above which the phase transition is strongly first order.}
	\label{fig:PTstrength}
\end{center}
\end{figure}
The EWPT strength $R \equiv v(T)/T$ can be computed both for $T_c$ and $T_n$, as shown in Fig.~\ref{fig:PTstrength}. Since the EWPT effectively starts at $T_n$, its strength is better estimated by $R_n$. A strongly first order EWPT, avoiding baryon number washout after electroweak baryogenesis, requires $R_n \gtrsim 1$ (see {\it e.g.} \cite{Moore:1998swa,Quiros:1999jp}), which for GMESB scenarios occurs already for $X \gtrsim 0.08$. The logarithmic term in (\ref{Higgs_potential}) coming from the breaking of scale invariance thus leads to a strong EWPT even for small deviations of $\lambda$ from its SM value. As $X$ grows, a large potential barrier between the minima develops. The amount of supercooling required to tunnel may then be large, and $T_n$ will be substantially lower than $T_c$, resulting in $R_n$ being significantly larger than $R_c$ (as shown in Fig.~\ref{fig:PTstrength}). For $X \gtrsim 0.47$ the symmetric vacuum is actually metastable (with a lifetime longer than the age of the Universe), 
leading to an inconsistent cosmology.

Once the EWPT starts, bubbles of true vacuum nucleate and expand, filling the entire Universe. The velocity $v_w$ of the expanding bubbles can be computed by solving a set of hydrodynamic equations~\cite{Moore:1995ua, Huber:2013kj}, since the plasma friction on the expanding bubble walls is known for the SM \cite{Moore:1995ua}. Stationary state bubbles expand either as subsonic {\it deflagrations} or as 
supersonic {\it detonations} (see {\it e.g.} \cite{Espinosa:2010hh}), the sound speed of a relativistic plasma being $c_s = 1/\sqrt{3} \sim 0.577$.
Subsonic bubbles could potentially lead to baryogenesis for a strong enough EWPT, $R_n \gtrsim 1$. Supersonic bubbles do not allow in general for baryogenesis (see however 
\cite{Caprini:2011uz}), but collisions of fast moving bubbles at the end of the EWPT can be a powerful source of gravitational waves \cite{Kamionkowski:1993fg,Caprini:2007xq,Huber:2008hg}. For very strong phase transitions the bubbles become ultrarelativistic and enter a ``{\it runaway}" (continuously accelerating) regime \cite{Bodeker:2009qy}, leading to very efficient gravitational wave production \cite{Espinosa:2010hh}. For the GMESB scenarios discussed here, we show in Fig.~\ref{fig:PTstrength} the ranges of $X$ for which {\it deflagrations}, {\it detonations} and {\it runaway} are realized. We see that $X\gtrsim 0.3$ leads to {\it runaway} bubbles, the best possible scenario for gravitational wave production.

\vspace{-5mm}

\subsection{III. Gravitational Wave Production}

\vspace{-3mm}

As discussed in the previous section, a large stochastic background of gravitational waves (GW) is expected for GMESB scenarios in a sizable portion of the allowed range of 
$X$, generated when bubbles collide at the end of the EWPT \cite{Kamionkowski:1993fg,Caprini:2007xq,Huber:2008hg,Grojean:2006bp,Child:2012qg}. The peak amplitude $h^2\Omega_{\rm peak}$ and peak frequency $f_{\rm peak}$ of such a GW background can be estimated to be \cite{Huber:2008hg}
\begin{small}
\be
	\label{GW_amplitude}
	h^2\Omega_{\rm peak} \simeq 10^{-6} \, 
	\left(\frac{H_{*}}{\beta}\right)^2\left(\frac{\kappa \, \alpha}{1+\alpha}\right)^2 
	\frac{1.84 \, v^3_w}{0.42 + v^2_w}
\ee 
\end{small}
\begin{small}
\be
	\label{GW_frequency}
	f_{\rm peak} \simeq 10^{-2} \mathrm{mHz} \,  
\left(\frac{T}{100 \, \mathrm{GeV}}\right)^2 \frac{\beta}{H_{*}} \frac{1.02}{1.8 + v^2_w}
\ee
\end{small}
where $\alpha$ is the ratio of vacuum energy to the energy stored in radiation, $\kappa$ is the efficiency in converting this vacuum energy into kinetic energy
that can lead to GW production \cite{Kamionkowski:1993fg,Espinosa:2010hh}, and $H_{*}/\beta$ roughly corresponds to the mean bubble size (normalized to the Hubble radius) at the time of collision. $\beta^{-1}$ gives also an estimate of the duration of the EWPT (taking the argument of the exponential in (\ref{nucleation_rate}) as a function of time,  expanding around the time $t_*$ at which the EWPT is completed and keeping only the leading order term \cite{Turner:1992tz} allows to write the nucleation probability as $P(t)\sim e^{\beta (t-t_*)}$). Note that all the previous discussion applies only when the EWPT proceeds much faster than the rate of expansion of the Universe ($\beta^{-1}/H_{*}^{-1} \ll 1$). Close to metastability $\beta$ can turn negative, and we have to use the procedure described in \cite{Huber:2008hg} to define the mean bubble size.

\begin{figure}[ht]
\begin{center}
	\includegraphics[width=0.45\textwidth, clip]{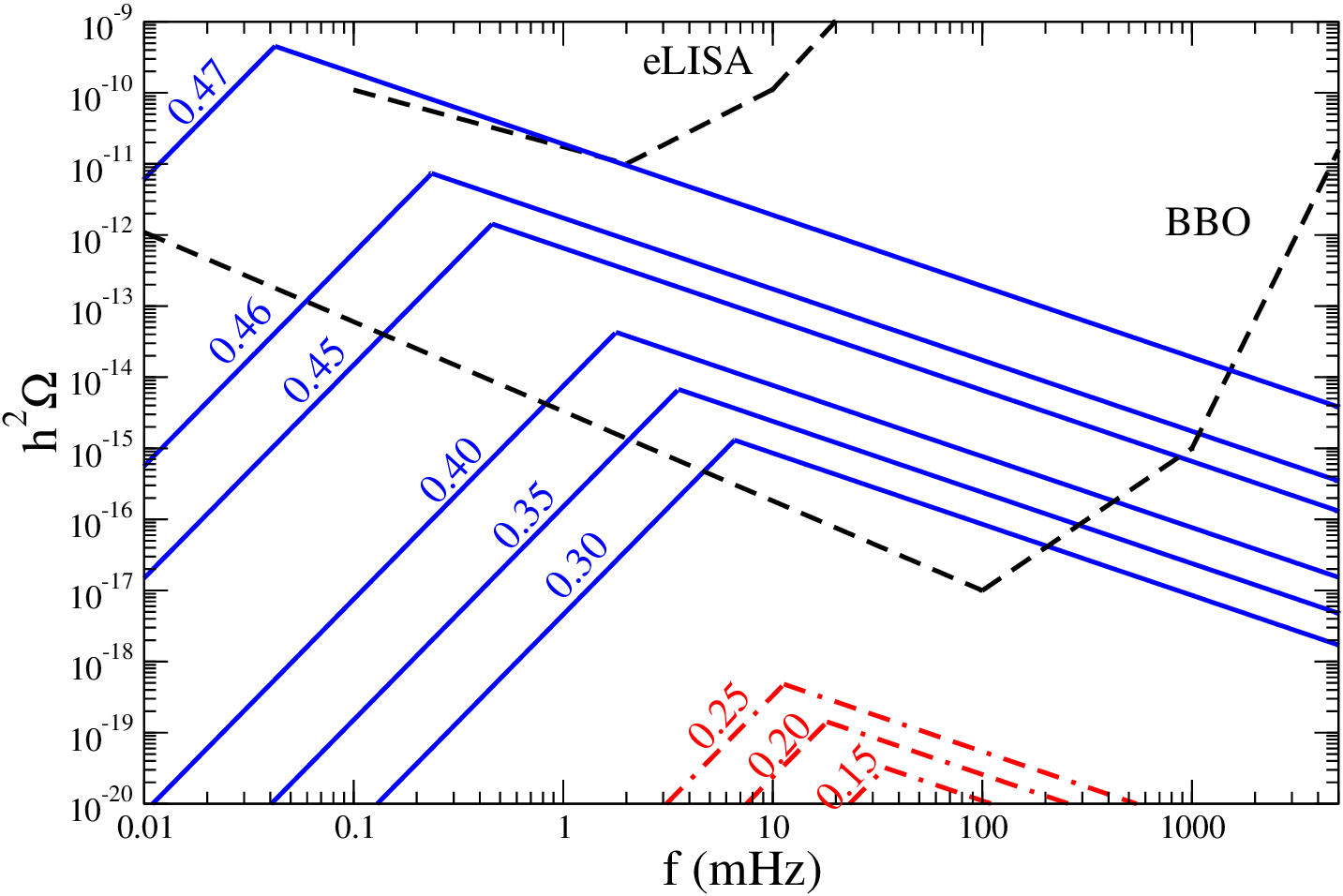}
	\caption{\small $h^2\Omega_{\mathrm{GW}}(f)$ for various values of $X$, as indicated in the red ({\it detonations}) and 
	blue ({\it runaway}) curves. The black dashed-lines are the sensitivity curves of eLISA and BBO.}
	\label{fig:GW}
	\end{center}
\end{figure}

For frequencies smaller than $f_{\rm peak}$ the GW spectrum grows as $f^{3}$ \cite{Caprini:2007xq}, whereas it falls off as $f^{-1}$ 
for large frequencies \cite{Huber:2008hg}. In Fig.~\ref{fig:GW} we show the GW spectrum for various values of $X$.   
For {\it detonations} (red spectra) the amplitude seems too small to be observable. For deflagrations the results are even smaller. Note however that recently plasma sound-waves formed upon bubble collisions have been identified as a strong source of GW \cite{Hindmarsh:2013xza}, possibly leading to an $H_{*}/\beta$ enhancement of the signal for {\it detonations} and  {\it deflagrations}.
For values of $X$ in the region $0.29 \lesssim X \lesssim 0.47$, correponding to {\it runaway} (blue spectra), the GW background would be observable by BBO, and could even be close to the sensitivity curve of eLISA (sound waves can be neglected in in this case). This could be a smoking-gun signature of GMESB scenarios in the absence of new physics at LHC. 

\vspace{-5mm}

\subsection{IV. Constraints on $f_c$ and the Higgs self-coupling}

\vspace{-3mm}

GMESB scenarios lead to modifications of the Higgs trilinear self-coupling $\lambda^{hhh}$. Expanding (\ref{Higgs_potential}) around $h=v$ (with $\bar{h}\equiv h-v$), we find
\be
	V_0 \supset \frac{m_h^2}{2}\, \bar{h}^2 + \frac{m_h^2}{6v}(3+2X)\, \bar{h}^3 + \frac{m_h^2}{24v^2}(3+4X)\, \bar{h}^4
\ee
so that $(\lambda^{hhh}/\lambda^{hhh}_{\mathrm{SM}}) -1 = 2X/3$. With an integrated luminosity of $3$~ab$^{-1}$, the planned upgrade of LHC would be able to measure deviations of $\lambda^{hhh}$ from its SM value with $30$~\% accuracy~\cite{Dawson:2013bba}, being only sensitive to $X > 0.45$, already on the edge of the allowed region of parameter space. While the planned linear $e^{+}\, e^{-}$ collider ILC would not improve this precision \cite{Baer:2013cma}, the $e^{+}\, e^{-}$ collider CLIC would be able to measure $\lambda^{hhh}$ with about $15\%$ accuracy \cite{Aicheler:2012bya}, probing the region in which {\it runaway} occurs. 

\vspace{1mm}

\begin{figure}[ht]
\begin{center}
	\includegraphics[width=0.45\textwidth, clip]{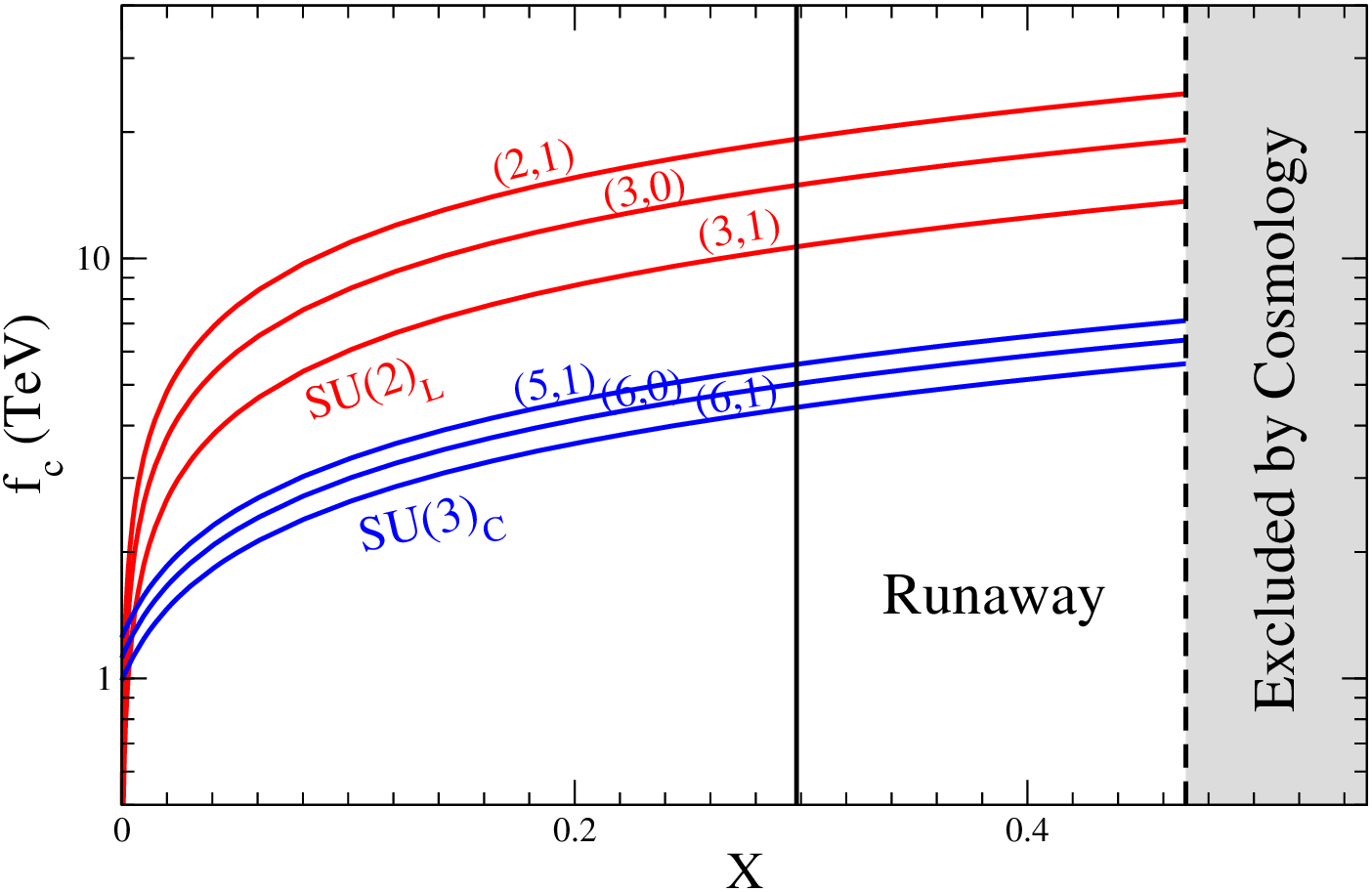}
	\caption{\small Upper bound on $f_{c(2)}$ (red) and $f_{c(3)}$ (blue) for various combinations of $(N_F^n, N_B^n)$.}
	\label{fig:fc}
	\end{center}
\end{figure}

It is also possible to obtain further information on how $X$ and the scale $f_c$ depend on the particle content of the hidden sector, 
by considering a (weakly coupled) GMESB scenario with the following generic assumptions: (i) particles in the hidden sector do not couple simultaneously to $SU(2)$ and $SU(3)$, which results in two independent hidden sectors, each with its own breaking scale $f_{c(n)}$ ($n=2,3$); (ii) bosons (fermions) belonging to the same hidden sector have a common anomalous dimension $\gamma_B^{(n)}$ ($\gamma_F^{(n)}$), which must be negative for consistency~\cite{Abel:2013mya}. As previously noted, the latter 
implies $X>0$, so that these GMESB scenarios naturally yield an EWPT significantly stronger than that of the SM. Requiring the 1-loop $\beta$-functions to vanish in the UV allows to write the scales $f_{c(n)}$ in terms of the number of bosons $N_B^{n}$ and fermions $N_F^{n}$ in each hidden sector, their anomalous dimensions, and $X$ (see \cite{Abel:2013mya} for details). In order to preserve perturbativity we require $-1\lesssim\gamma^{(n)}_{B,F}<0$, resulting in lower and upper bounds on the combination $N_{B}^{n}+2N_F^{n}$. This also results in an upper bound on $f_{c(n)}$ (the bound is automatically saturated in the case of purely fermionic hidden sectors) as a function of $X$, which decreases with increasing $N_{B}^{n}+2N_F^{n}$, as shown in Fig.~\ref{fig:fc} for various allowed $(N_F^n, N_B^n)$ combinations. 
The cosmological bounds on $X$ derived in previous sections then impose upper bounds on the scale of  breaking of scale invariance in the hidden sectors, namely $f_{c(2)}\lesssim 25$~TeV and $f_{c(3)}\lesssim 7$~TeV. In GMESB scenarios with a common breaking scale $f_{c(2)} = f_{c(3)} = f_{c}$, 
Fig. \ref{fig:fc} implies $X \ll 1$ and $f_{c} \sim 1$ TeV, within LHC reach. In contrast, for scenarios with a significant hierarchy between $f_{c(2)}$ and $f_{c(3)}$, 
large values of $X$ are preferred, together with both $f_{c(2)}, f_{c(3)} \gtrsim 3$ TeV, making the hidden sectors hard to be probed directly at LHC. 
This highlights the fact that a strong EWPT is anticorrelated with the LHC search prospects for GMESB scenarios.

\vspace{-5mm}

\subsection{V. Conclusions}

\vspace{-3mm}
We have shown that models with gauge mediation of breaking of scale invariance (GMESB scenarios) generically lead to a strong electroweak phase transition. Such setups could easily escape direct detection at LHC,  especially if GMESB is dominated by $SU(2)_\mathrm{L}$ gauge interactions, where the breaking scale $f_c \gtrsim 5$ TeV.
The measurement of a nonstandard cubic Higgs coupling by CLIC combined with the most interesting observation of a large primordial gravitational wave background is a promising route for testing these GMESB scenarios.


\vspace{3mm}

\begin{acknowledgments}
We thank A. Mariotti and S. Abel for very useful discussions and comments on the manuscript.
S.H. and J.M.N. are supported by the Science Technology and Facilities
Council (STFC) under grant No. ST/J000477/1. 
G.C.D. is supported by CAPES (Brazil) under grant No. 0963/13-5.
\end{acknowledgments}


\begin{thebibliography}{99}
\vspace*{-1cm}
%
\bibitem{Aad:2012tfa}
  G.~Aad {\it et al.}  [ATLAS Collaboration],
  Phys.\ Lett.\ B {\bf 716} (2012) 1
  [arXiv:1207.7214 [hep-ex]].

\bibitem{Chatrchyan:2012ufa}
  S.~Chatrchyan {\it et al.}  [CMS Collaboration],
  Phys.\ Lett.\ B {\bf 716} (2012) 30
  [arXiv:1207.7235 [hep-ex]].

\bibitem{Salam:1970qk}
A.~Salam and J.~A.~Strathdee,
Phys.\ Rev.\  {\bf 184} (1969) 1760.

\bibitem{Bardeen:1995kv}
  W.~A.~Bardeen,
  FERMILAB-CONF-95-391-T.

\bibitem{Tavares:2013dga}
  G.~Marques Tavares, M.~Schmaltz and W.~Skiba,
  Phys.\ Rev.\ D {\bf 89} (2014) 015009
  [arXiv:1308.0025 [hep-ph]].
  
\bibitem{Tamarit:2013vda}
  C.~Tamarit,
  JHEP {\bf 1312} (2013) 098
  [arXiv:1309.0913 [hep-th]].
  
  \bibitem{Antipin:2013exa}
  O.~Antipin, M.~Mojaza and F.~Sannino,
  arXiv:1310.0957 [hep-ph].  

\bibitem{Abel:2013mya}
  S.~Abel and A.~Mariotti,
  arXiv:1312.5335 [hep-ph].
  
 \bibitem{Litim:2011cp}
  D.~F.~Litim,
 Phil.\ Trans.\ Roy.\ Soc.\ Lond.\ A {\bf 369} (2011) 2759
 [arXiv:1102.4624 [hep-th]]. 

\bibitem{Kajantie:1996mn}
  K.~Kajantie, M.~Laine, K.~Rummukainen and M.~E.~Shaposhnikov,
  Phys.\ Rev.\ Lett.\  {\bf 77} (1996) 2887
  [hep-ph/9605288].

\bibitem{Karsch:1996yh}
  F.~Karsch, T.~Neuhaus, A.~Patkos and J.~Rank,
  Nucl.\ Phys.\ Proc.\ Suppl.\  {\bf 53} (1997) 623
  [hep-lat/9608087].

\bibitem{Csikor:1998eu}
  F.~Csikor, Z.~Fodor and J.~Heitger,
  Phys.\ Rev.\ Lett.\  {\bf 82} (1999) 21
  [hep-ph/9809291].

\bibitem{Dolan:1973qd}
  L.~Dolan and R.~Jackiw,
  Phys.\ Rev.\ D {\bf 9} (1974) 3320.
  
  \bibitem{Carrington:1991hz}
  M.~E.~Carrington,
  Phys.\ Rev.\ D {\bf 45} (1992) 2933.

  \bibitem{Coleman:1977py}
  S.~R.~Coleman,
  Phys.\ Rev.\ D {\bf 15} (1977) 2929
   [Erratum-ibid.\ D {\bf 16} (1977) 1248].
   
   \bibitem{Linde:1980tt}
  A.~D.~Linde,
  Phys.\ Lett.\ B {\bf 100} (1981) 37;
  Nucl.\ Phys.\ B {\bf 216} (1983) 421
   [Erratum-ibid.\ B {\bf 223} (1983) 544].


\bibitem{Moore:1998swa}
  G.~D.~Moore,
  Phys.\ Rev.\ D {\bf 59} (1999) 014503
  [hep-ph/9805264].

   \bibitem{Quiros:1999jp}
  M.~Quiros,
  hep-ph/9901312.
      
 \bibitem{Moore:1995ua}
  G.~D.~Moore and T.~Prokopec,
  Phys.\ Rev.\ Lett.\  {\bf 75} (1995) 777
  [hep-ph/9503296]; Phys.\ Rev.\ D {\bf 52} (1995) 7182
  [hep-ph/9506475].
 
\bibitem{Huber:2013kj}
  S.~J.~Huber and M.~Sopena,
  arXiv:1302.1044 [hep-ph].
   
   
 \bibitem{Espinosa:2010hh}
  J.~R.~Espinosa, T.~Konstandin, J.~M.~No and G.~Servant,
  JCAP {\bf 1006} (2010) 028
  [arXiv:1004.4187 [hep-ph]].  
   
 \bibitem{Caprini:2011uz}
  C.~Caprini and J.~M.~No,
  JCAP {\bf 1201} (2012) 031
  [arXiv:1111.1726 [hep-ph]].  
   
 \bibitem{Kamionkowski:1993fg}
  M.~Kamionkowski, A.~Kosowsky and M.~S.~Turner,
  Phys.\ Rev.\ D {\bf 49} (1994) 2837
  [astro-ph/9310044].
  
  \bibitem{Caprini:2007xq}
  C.~Caprini, R.~Durrer and G.~Servant,
  Phys.\ Rev.\ D {\bf 77} (2008) 124015
  [arXiv:0711.2593 [astro-ph]].
  
  \bibitem{Huber:2008hg}
  S.~J.~Huber and T.~Konstandin,
  JCAP {\bf 0809} (2008) 022
  [arXiv:0806.1828 [hep-ph]].  
  
  \bibitem{Child:2012qg}
  H.~L.~Child and J.~T.~Giblin, Jr.,
  JCAP {\bf 1210} (2012) 001
  [arXiv:1207.6408 [astro-ph.CO]].
  
 \bibitem{Hindmarsh:2013xza}
  M.~Hindmarsh, S.~J.~Huber, K.~Rummukainen and D.~J.~Weir,
  Phys.\ Rev.\ Lett.\  {\bf 112} (2014) 041301
  [arXiv:1304.2433 [hep-ph]].
   
 \bibitem{Bodeker:2009qy}
  D.~Bodeker and G.~D.~Moore,
  JCAP {\bf 0905} (2009) 009
  [arXiv:0903.4099 [hep-ph]].
  
 \bibitem{Grojean:2006bp}
  C.~Grojean and G.~Servant,
  Phys.\ Rev.\ D {\bf 75} (2007) 043507
  [hep-ph/0607107]. 
  
\bibitem{Turner:1992tz}
  M.~S.~Turner, E.~J.~Weinberg and L.~M.~Widrow,
  Phys.\ Rev.\ D {\bf 46} (1992) 2384.

 \bibitem{Dawson:2013bba}
  S.~Dawson {\it et al.},
  arXiv:1310.8361 [hep-ex].
  
 \bibitem{Baer:2013cma}
  H.~Baer {\it et al.},
  arXiv:1306.6352 [hep-ph]. 
  
 \bibitem{Aicheler:2012bya}
  M. Aicheler {\it et al.},
  CERN-2012-007. 
  
\end{thebibliography}
\end{document}